# Field manipulation of Weyl modes in an ideal Dirac semimetal

Jingyuan Zhong[1,#], Jianfeng Wang[1,#,*], Ming Yang[1,2], Jie Liu[3], Zhizhen Ren[1], Anping Huang[1], Zhixiang Shi[4], Zengwei Zhu[3], Yan Shi[5], Weichang Hao[1,*], Jincheng Zhuang[1,*], Yi Du[1,2,*]

[1] School of Physics, Beihang University, Haidian District, Beijing 100191, People's Republic of China
[2] The Analysis & Testing Center, Beihang University, Beijing 100191, People's Republic of China
[3] Wuhan National High Magnetic Field Center and School of Physics, Huazhong University of Science and Technology, Wuhan, 430074, People's Republic of China
[4] School of Physics and Key Laboratory of the Ministry of Education, Southeast University, Nanjing 211189, People's Republic of China
[5] School of Automation Science and Electrical Engineering, Beihang University, Beijing 100191, People's Republic of China

[#]Jingyuan Zhong and Jianfeng Wang contributed equally to this work.
*Correspondence authors. E-mail: jincheng@buaa.edu.cn; wangjf06@buaa.edu.cn; whao@buaa.edu.cn; yi_du@buaa.edu.cn

**Abstract**

**The emergent Weyl modes with the broken time-reversal symmetry or inversion symmetry provide large Berry curvature and chirality to carriers, offering the realistic platforms to explore topology of electrons in three-dimensional systems. However, the reversal transition between different types of Weyl modes in a single material, which is of particular interest in the fundamental research in Weyl physics and potential application in spintronics, is scarcely achieved due to restriction of inborn symmetry in crystals. Here, by tuning the direction and strength of magnetic field in an ideal Dirac semimetal, $Bi_4(Br_{0.27}I_{0.73})_4$, we report the realization of multiple Weyl modes, including gapped Weyl mode, Weyl nodal ring, and coupled Weyl mode by the magnetoresistivity measurements and electronic structure calculations. Specifically, under a magnetic field with broken mirror symmetry, anomalous Hall effect with step feature results from the large Berry curvature for the gapped Weyl mode. A prominent negative magnetoresistivity is observed at low magnetic field with preserved mirror symmetry and disappears at high magnetic field, which is correlated to the chiral anomaly and its annihilation of Weyl nodal ring, respectively. Our findings reveal distinct Weyl modes under the intertwined crystal symmetry and time-reversal breaking, laying the foundation of manipulating multiple Weyl modes in chiral spintronic network.**

## Introduction

Symmetry and topology are symbiotic in condensed matter physics, and their interplay is a fundamental aspect classifying the different topological phases such as topological insulator, topological nodal point/line/surface semimetal, and topological crystalline insulator, etc. [1-5] The topological phase transitions are expected to be generated when the symmetry of a system is broken by varying the ordering parameters or introducing the external perturbation. For example, Weyl physics can be induced by breaking inversion symmetry (IS) or time-reversal symmetry (TRS) in a Dirac semimetal, along with the abundant intriguing phenomena like anomalous Hall effect (AHE), unconventional thermal magnetic responses, negative magnetoresistivity (NMR) by chiral anomaly, and three-dimensional (3D) quantum Hall effect. [6-13] Despite numerous Weyl materials revealed in non-centrosymmetric or magnetic systems so far,[10-14] a symmetry guideline for searching for various Weyl modes and designing their evolution is still unrevealed, which is crucial for the manipulation of Weyl fermion towards the chiral device application.

Among various modulation ways, the magnetic field is regarded as an effective and reversible method to evoke topological phases transitions, which has been theoretically proposed in the 3D warped semimetals [15] and experimentally verified in $HfTe_5$ (from topological insulator to quasi-one-dimensional (quasi-1D) gapped Weyl mode). [10,14,16-17] Here, we propose a strategy to realize the magnetic field manipulation of various Weyl modes from a Dirac semimetal (DSM). As shown in Fig. 1**a**, a DSM with IS and TRS possesses the band structure with four-fold degenerate Dirac point (DP). Introducing magnetic field $B$ can break TRS and lift the spin degeneracy, engendering the split of Dirac bands into two Weyl bands with opposite spins. The chiral symmetry for Weyl fermions can be also dictated by the crystalline symmetries. If $B$ breaks the symmetry maintaining the chiral feature, the disturbance of spin-orbital coupling (SOC) leads to a gapped Weyl mode with the energy gap $\Delta_W$ as shown in Fig. 1**b**. When $B$ preserves some crystal symmetry, *e.g.*, along the normal direction of a mirror plane $M$, the gapless Weyl points (WPs) or Weyl nodal ring (WNR) can be formed, as displayed in Fig. 1**c**. Noticeably, magnetic field tends to circulate and localize electrons in real space, which broadens the spreading of wavefunctions (denoted by the grey peaks) in momentum space. [18-20] These wavefunctions can overlap with each other in momentum space when the $B$ is large enough. Consequently, the chiral charges are able to tunnel between chirality-opposite WPs to open the magnetic tunneling gap $\Delta_T$ without breaking chiral symmetry in Fig. 1**d**. As a result, the magnetic field manipulation of various Weyl modes can be realized based on this symmetry guideline. Nevertheless, these topological phase transitions can not be directly measured by angle-resolved photoemission spectroscopy (ARPES) due to the influence of magnetic field. As an alternate, the magnetoresistivity measurement has been applied to identify the different Weyl modes in previous reports. [18,20] Although the previous work reports the realization of Weyl modes in various materials, their transport behaviors are unavoidably disturbed by irrelevant, conventional electrons [4,7,10-14], hindering the observation of the multiple Weyl modes in a single material. Therefore, searching for an ideal DSM with only Dirac bands near Fermi surface to exclude the contribution from other bands in transport measurements is highly required to experimentally explore the chiral physics.

In this work, we identify an ideal van der Waals DSM material Bi$_4$(Br$_{0.27}$I$_{0.73}$)$_4$ with only Dirac bands residing in the Fermi level, which is a topological phase transition critical point bridging higher-order topological insulating phase and dual topological order. [21] The magnetoresistivity, Hall measurements, and first-principles calculations are performed to reveal its electronic structures under a magnetic field along different directions. The AHE is observed when the magnetic field is perpendicular to (001) plane, which is correlated to the large Berry curvature in the gapped Weyl mode. Shubnikov–de Haas (SdH) oscillation and consequent NMR are explored with the magnetic field along chain direction, which are evoked by splitting of Landau level and chiral anomaly in the WNR, respectively. Continuously enhancing the magnetic field upon 21.5 T, the breakdown of chiral anomaly is observed, indicating the annihilation of the WPs. All the experimental results are supported by the calculations, and finally, a phase diagram of Weyl physics as the function of magnetic field and temperature is established, constructing the various Weyl modes into a unified picture in a single crystal quantum system and establishing the close relationship between the crystal symmetry and chiral electrons.

**Results**

**Ideal DSM in Bi$_4$(Br$_{0.27}$I$_{0.73}$)$_4$.** Bi$_4$X$_4$ ($X$ = Br, I) is assembled with 1D bismuth skeleton surrounded by halide atoms under the van der Waals (vdW) force, possessing IS and a single mirror plane $M_y$ (marked by black dashed line) perpendicular to the chain direction as shown in Fig.2**a**. The bulk Brillouin zone (BZ) with (001) surface projected BZ is shown in Fig. 2**b**. This system possesses abundant topological phases, including weak topological insulator, high-order topological insulator, dual topological phase, and topological trivial insulator, by controlling mole ratio of Br and I. [21-35] Having the same stacking structure as Bi$_4$Br$_4$, a Dirac semimetal phase can be realized at the critical point of around 73% I concentration (Fig. S1 and Fig. S2), which has been demonstrated by the Dirac cones (labelled by black arrow) in both of the ARPES spectra in Fig. 2**c** and density functional theory (DFT) calculation with virtual crystal approximation (VCA) method in Fig. 2**d** and Fig. S3. Astonishingly, only the bands originated from the Bi $p$ orbits locate near the Fermi surface, evidencing Bi$_4$(Br$_{0.27}$I$_{0.73}$)$_4$ can be an ideal material platform to manipulate and investigate the multiple Weyl modes proposed in Fig. 1 (the other band residing in Fermi level will be pushed away to high energy position after introducing magnetic field). It is noticeable that the halide vacancies induce semiconducting-like transport behavior in pristine Bi$_4$Br$_4$. [36] This localization effect becomes negligible in Bi$_4$(Br$_{0.27}$I$_{0.73}$)$_4$ due to the plenty of iterant carriers in Dirac semimetal Bi$_4$(Br$_{0.27}$I$_{0.73}$)$_4$.

**AHE and large Berry curvature evoked by the gapped Weyl mode.** The typical six-wire method is adopted to measure the magnetoresistivity and Hall resistivity of Bi$_4$(Br$_{0.27}$I$_{0.73}$)$_4$, as shown in the schematic in Fig. 3**a**, where the current flows through chain direction (lattice $b$ axis) and Hall voltage is measured along lattice $a$ axis. Measurement is performed with the varied angles $\varphi$ between magnetic field and $M_y$ (insets of Fig. 3**b**). The magnetic field-dependent anomalous Hall resistivity $\rho_{yx}^A$ for S1 in Fig. 3**b** shows the step-like shape with two stages instead of the linear or curved dispersion in a large $\varphi$ region ranging from 0° to 60°, which is a ubiquitous signal also identified in other crystals from different batches (see Fig. S4 in Supplementary Information). Furthermore, the $\rho_{yx}^A$ results as a function of normal magnetic field component ($B\cos\varphi$) at different $\varphi$ collapse into a single curve, implying the signal of $\rho_{yx}^A$ is mainly contributed by $B_z$ component of magnetic field. The two band model is applied to fit the Hall signals, where the large deviation between the fitting lines and experimental results excludes the possibility of multiple band effect (see Fig. S5 for multiple band fitting results for different samples). The stage-like behavior in Hall resistivity surviving at high temperature (up to 300 K) and strong magnetic field (up to 57 T) (see Fig. S6), and is similar with the AHE in the magnetic system, where the stage is correlated to the saturated magnetic moment. [37] Nevertheless, there are no magnetic atoms and domains in this quasi-1D system, excluding the contribution of magnetic domains to the AHE (See Fig. S7 and Fig. S8 in detail). The quantum geometry of the electronic bands can also evoke the nonlinear Hall effect, where only the Berry curvature, which acts like an effective magnetic field in momentum space and conforms to the condition of the reserved inversion symmetry in $Bi_4(Br_{0.27}I_{0.73})_4$, contribute the linear AHE. [7,37-46] The contribution of non-zero Berry curvature to Hall resistivity can be classified into two parts, the scattering independent part (intrinsic and side-jump contribution) and scattering dependent part (localized hopping and skew scattering contribution). [37-39] In this case, the saturated Hall conductivity $\sigma_{xy}^A$ can be expressed as:

$$\sigma_{xy}^A = \alpha \sigma_{yy}^{1.6} + \sigma_{xy}^{Ind.}, \tag{1}$$

where $\alpha$ is a constant correlated with residual resistivity, $\sigma_{yy}$ is the conductivity at zero field, and $\sigma_{xy}^{Ind.}$ is the anomalous Hall conductivity (AHC) correlated to the scattering independent part. Fig. 3c exhibits the $\sigma_{xy}^A$ values as a function of $\sigma_{yy}$ measured at different temperatures, where all the points can be well fitted by Eq. (1) with a finite interception value. The validity of our fitting results is supported by the comparative study of fitting lines by using different exponents and intercepts, as shown in Fig. S9 in Supplementary Information. The 1.6 exponent indicates existence of the localized hopping conduction in dirty metals (in which conductivity less than 10000 S/cm), [40-42] and the nonzero interception value (~ 169 S/cm) is a direct reflection of the scattering independent contribution, $\sigma_{xy}^{Ind.}$. It is noted that both extrinsic side-jump mechanism and intrinsic contribution are supposed to be scattering independent, which are not distinguishable here from scattering plot. [37,40-42] Consequently, the well fitted experimental data in Fig. 3c by Eq. (1) indicates that the step-like feature originates from both scattering dependent localized hopping effect and scattering independent contribution, where non-zero Berry curvature behaves as the source to deflect electrons. [41] The anomalous Hall angle (AHA) standing for the deflection angle induced by Berry curvature can be calculated by $\tan^{-1}(\frac{\sigma_{xy}^A}{\sigma_{yy}})$, and is 20.8° at $T$ = 20 K (the inset of Fig. 3c). The large AHA in $Bi_4(Br_{0.27}I_{0.73})_4$ is contributed by both localized hopping conduction and scattering independent contribution of Berry curvature (see Supplementary Note 7 for detailed illustration of Berry curvature contribution to carrier deflection), which is reasonable to emerge in strongly disordered metal (*i.e.* bad metal with low conductivity) as shown in Fig. S10, coinciding with the high I dopants concentration (more disorders than pure $Bi_4Br_4$ sample). It should be noted that Berry curvature can be characterized by scanning-superconducting quantum interference device or reflectance magnetic circular dichroism, which requires further study in the future. [47-50]

The calculated band structure and intrinsic AHC correlated with Berry curvature of $Bi_4(Br_{0.27}I_{0.73})_4$ under an effective magnetic field $B_z$ = 2 T and $B_z$ = 4 T (see method for estimation of effective magnetic field) are shown in Fig. 3**d** and Fig. 3**e**, respectively. In our calculations, the external magnetic field is applied in a Zeeman-like manner (see calculation methods). Although the Landau level effect is not included in the simulations, the basic physical results induced by the Berry curvature and gapless Weyl modes can be definitely determined. With the out-of-plane magnetic field breaking mirror symmetry, the gapped Weyl modes are observed, which are consistent with the schematic in Fig.1**b**. Fundamentally, the relationship between $\sigma_{xy}^A$ and $\Omega_z$ can be written as: $\sigma_{xy}^A \propto \int_{BZ}^{BZ} \Omega_z dk$ where $\Omega_z = \sum_{n \in occupied} \Omega_z^n$ coinciding with the emergent large AHC near gaps in Fig. 3**d** and Fig. 3**e**. [7,37] It should be noted that there is a *n*-type doping effect in this quasi-1D system due to the vacancies of halide elements, [36] resulting in the DP locating at 0.14 eV below $E_F$ at zero magnetic field (Fig. 2**c**). With increasing magnetic field, the energy difference between $E_F$ and DP becomes small due to the increased degeneracy of Landau bands. By comparing the calculated AHC and experimental results, we attribute the small stage I to the Landau band surpassing $E_F$, corresponding to small and non-saturated $\Omega_z$, as shown in Fig. 3**d**. On the contrary, stage II is robust, indicating the large and saturated $\Omega_z$ after $B_z$ = 4 T, which is confirmed by the calculated AHC at $E_F$ in Fig. 3**e** (see Fig. S11 for details). The $\Omega_z$ distributions summed to $E_F$ in Fig. 3**d** and Fig. 3**e** are shown in Fig. 3**f**, where non-saturated $\Omega_z$ with a hollow circle and saturated $\Omega_z$ with a solid circle are observed for $B_z$ = 2 T and $B_z$ = 4 T respectively. The effect of *n*-type doping level on the AHE is discussed in Supplementary Note 10 with Fig. S12.

**NMR phenomenon and the formation of WNR.** The longitudinal resistivity measurements under magnetic field $\rho_{yy}(B)$ are also performed to characterize the transport properties and to infer underlying topological features. Fig. 4**a** shows three $\rho_{yy}(B)$ curves with different magnetic field direction: along lattice *a* axis, perpendicular to (001) surface, and along lattice *b* axis. The curve shapes are almost the same for the first two cases, which is made up by a cusp shape induced by a weak antilocalization at low field and positive magnetoresistivity at high field. [24-26] Interestingly, the curve shape is different when magnetic field applied along lattice *b* axis. Three oscillation peaks emerge after the disappearance of cusp with the increment of magnetic field. The analysis of this oscillation in Fig. 4**b** exhibits the equal 1/*B* interval, demonstrating that the typical Shubnikov–de Haas (SdH) oscillation with oscillation frequency $B_F$ around 2.94 T is formed by the Landau level splitting with the magnetic field. [51] Moreover, the Landau bands index shows an intercept value within ± 1/8, implying nontrivial Berry phase of charge carriers and that the system enters into ultraquantum limit at ~ 2.6 T. [12,52] It is noticeable that, in another quasi-1D topological crystal ZrTe$_5$, electrons enter ultraquantum limit at a compatible field (~ 1.3 T). [13] The 0.5 intercept correlated to nontrivial Berry phase has also been reported. [43] It is due to the different definition methods of integer Landau index compared to our work. Our results are consistent with these references. The non-trivial Berry phase coincides with the Dirac fermions near the Fermi surface in Bi$_4$(Br$_{0.27}$I$_{0.73}$)$_4$. Prominent SdH oscillation only appears under $B \parallel b$ configuration, which can be explained by the small effective mass and high mobility of charge carriers in the gapless linear band structure along chain direction compared to other two directions. It is noticeable that the quasi-1D in a closed Fermi surface feature allows electron hopping to produce SdH oscillations under magnetic field. The NMR emerges after the termination of the

SdH oscillation as itinerant carriers are condensed into the lowest Landau bands, which is supported by the similar values between the estimated onset magnetic field of ultraquantum limit and transition field in Fig. **4a** and Fig. **4b** (see Supplementary Note 12 for details)). [8-9,51-53]. Detailed angle-dependent $\rho_{yy}$ curves under different magnetic field are shown in Fig. S13, indicating the narrow angular window for NMR. It should be noted that the charge puddle effect can induce isotropic (*i.e.* direction-independent) NMR phenomenon. [54] Such a possibility is excluded by the anisotropic NMR in Fig. **4a** and Kelvin probe force microscopy (KPFM) measurements in Fig. S7 and Fig. S8. In order to figure out the origin of the NMR, the band structure under the magnetic field along chain (*y*) direction is calculated. Fig. **4c** shows the calculated $E$ - $k_x$ - $k_z$ 3D band structure with an anisotropic WNR locating in $k_x$ - $k_z$ plane under $B_y$ = 4 T. The calculated energy momentum dispersions are exhibited in Fig. **4d**, where two special pairs of WPs in WNR are labelled as WP1 and WP2 along $k_x$ and $k_y$ directions, respectively. The band structure at $M$ along magnetic field $k_y$ direction possesses an energy gap. The WNR is composed of infinite pairs of WPs, and every two WPs with opposite chirality of one pair are located on opposite sides symmetric about the center point. However, due to the quasi-1D nature of the band, the WNR exhibits an elongated distribution, with either side maintaining approximately the same chirality, *i.e.* WP1+ or WP1-, except for the two endpoints (WP2+ and WP2-), as displayed in Fig. **4c**. Consequently, the WNR can be disassembled into infinite WP pairs, and its transport behavior falls into the same theoretical region of WPs. [19,55] In fact, NMR has been previously observed in the $\rho_{yy}$ measurements of Dirac semimetals and Weyl semimetals, where an additional transport channel along current direction for charge carrier connecting two WPs is provided to enhance the conductivity under the magnetic field. [7-11,19] In these cases, the magnetic field is with the same direction connecting two WPs (*i.e.* $B \parallel \Delta k_W$). Nevertheless, it is highlighted that when magnetic field is perpendicular to $\Delta k_W$ (*i.e.* $B \perp \Delta k_W$), NMR can still be induced by chiral anomaly as depicted in Fig. **4e**, where the electric field lifts chemical potential degeneracy ($\mu_- > \mu_+$) between two Weyl branches with opposite chirality as predicted in theory. [19] Consequently, chiral carriers with group velocity along crystal *b* axis are pumped and relaxed between WPs, offering additional chiral charge channels with enhanced conductivity along chain direction. Our $\rho_{yy}(B)$ results and calculations agree with each other, demonstrating the formation of WNR by $M_y$ protection as depicted in Fig. **1c** and the consequent NMR with the charge carriers pushed to lowest Landau level under the magnetic field parallel to current direction.

**Annihilation of Weyl node under the high magnetic field.** The $\sigma_{yy}$ curves of Bi$_4$(Br$_{0.27}$I$_{0.73}$)$_4$ sample are measured under a pulse magnetic field up to 57 T at different temperatures to investigate its transport behavior, as shown in Fig. **5a** and Fig. S14. For temperature below 50 K, the NMR data between two transition points, $B_0$ and $B_1$, can be nicely fitted by parabolic line. The NMR effect in ultraquantum limit can be described by: [8-9,11-12]

$$\sigma_{yy} = \sigma_0 + c_w B = \sigma_0 + \frac{e^3 \tau_a}{4\pi^2 \hbar^2 c} B, \qquad (2)$$

where $\sigma_{yy}$ is the longitudinal conductivity, $\sigma_0$ is residual conductivity, $c_w$ is the coefficient term, $e$ is elemental charge, $\tau_a$ is inter-node relaxation time, $\hbar$ is the reduced Planck constant, and $c$ is the light speed. The linear relationship between $\sigma_{yy}$ and $B$ can be inferred in the condition that the leading coefficient is a constant. [8-9] Since our results in Fig. **4b** and

Supplementary Note 12 indicate that the NMR region is correlated to the lowest occupied Landau bands, we believe that this inconsistency may be evoked by the magnetic field-dependent $\tau_a$. This hypothesis needs the further investigations in the future as the NMR tendency with magnetic field in ultraquantum limit region is still an open-ended question. The cusp at low field is correlated to the weak antilocalization effect, and parabolic shape corresponding to the NMR emerges at $B_0$. $\sigma_{yy}$ starts to diminish at $B_1$, indicating the vanish of the contribution of chiral anomaly.

In fact, the high magnetic field perpendicular to the connecting direction of WPs can induce magnetic tunneling between WPs with the finite coupling gap at original location of WPs, which can be modelled by: [20]

$$\sigma_{yy} = \sigma_0 + \sigma_1 e^{-\frac{\Delta_T(B)}{k_B T}}, \tag{3}$$

where $\sigma_1$ is conductivity from the lowest Landau band, $k_B$ is Boltzmann constant, $\Delta_T(B)$ denotes field-dependent magnetic-tunnelling gap. The Eq. (3) is applied to fit the region $B > B_1$ giving highly identical reproduction of experimental results. The fitted gap sizes of $\Delta_T(B)$ at different temperature are shown in Fig. 5b, which is compared to the thermal perturbation energy $k_B T$. The comparable energy values between $\Delta_T$ and $k_B T$ are reasonable to the opened gap case since the gap size need to be large enough to conquer thermal broadening effect. To achieve magnetic tunnelling of WPs, the distance between coupled WPs ($\Delta k_W$) should be comparable to the momentum expansion of DOS in WPs as the schematic in Fig. 1d. [18] It is reported that the inverse magnetic length has the direct relationship with the momentum expansion of DOS $1/l_B = \sqrt{eB/\hbar} \propto \Delta k_W$ [18-20] The estimated $\Delta k_W$ is 0.018 Å$^{-1}$ at 4.2 K (Fig. 5b) after taking $B_1 = 21.5$ T into account. The evolution of WNR and the distances between WP2 at $B_y = 4$ T and $B_y = 22$ T are shown in Fig. 5c, where $\Delta k_W$ value between WP2+ in the first BZ and WP2- in the second BZ is also 0.018 Å$^{-1}$ for $B_y = 22$ T, implying the complete chirality breakdown at $B_1$ induced by magnetic tunnelling between WP2 across the BZ boundary. In fact, the WPs with distance smaller than 0.018 Å$^{-1}$ (i.e. two WP1 separated along $k_x$) should be annihilated at $B_y = 4$ T indicating that the chiral anomaly discussed above mainly comes from WP2 as shown in Fig. 4e. Calculated $E - k_y$ dispersions near WP2 at $B_y = 4$ T and $B_y = 22$ T are presented in Fig. 5d and Fig. 5e, respectively, where magnetic tunnelling effect is absent (present) due to discrete (overlapped) DOS along $k_z$. Fig. 5f shows schematics of magnetic tunnelling effect, where overlapped DOS makes the lowest Landau bands couple with each other to form a gap $\Delta_T$, leading to chirality breakdown after $B_1$. Based on the above results and analysis, we plot a phase diagram as a function of temperature and magnetic field in Fig. 5g, where chiral anomaly and chirality breakdown, corresponding to the WNR and magnetic tunneling effect, respectively, can be reversibly modulated by the magnetic field.

To summarize, different Weyl modes have been modulated and identified in transport measurements by applying magnetic field, evidencing the successful engineering of highly pure Dirac or Weyl bands in bismuth halide Bi$_4$(Br$_{0.27}$I$_{0.73}$)$_4$ and the close relationship between crystal symmetry and chiral electrons. The exploration of these Weyl modes accompanied with the AHE is beneficial to the room-temperature device application of non-volatile storage and magnetic field detection. Our work simultaneously identifies the exotic phenomena, such as AHE, NMR, and annihilation of Weyl node, in one system, distinguishing from the individual exploration in other work. [4,7,10-12] The results reported in our work have been systematically validated through the excellent coincidence among the symmetry-based framework for designing Weyl states, magnetotransport data and fitting results, and first-principles simulations as well as the calculated parameters, which is rarely a task in previous reports. Furthermore, as the ideal Dirac semimetal with only the Dirac bands residing in the Fermi level, $Bi_4(Br_{0.27}I_{0.73})_4$ provides a material platform to realize more intriguing quantum transport behaviors contributed by quantum geometry after the symmetry breakdown. [43-46] The nondestructive manipulation of band topology with emergent Weyl modes, chiral anomaly and its annihilation, not only inspires deeper understanding towards Weyl physics, but also promotes the application of chiral spintronic devices.

**Methods**

**Crystal synthesization**. Solid-state reaction and the chemical vapor transport method were applied to grow the $Bi_4(Br_{0.27}I_{0.73})_4$ single crystals. Highly pure Bi, $BiBr_3$, and $BiI_3$ powders were mixed under Ar atmosphere in a glove box and sealed in a quartz tube under vacuum, where the mole ratio of $BiI_3$ to $BiBr_3$ were 7 : 3. The mixture was placed in a two-heating-zone furnace with the temperature gradient from 558 K to 461 K for 72 h. Crystal $Bi_4(Br_{0.27}I_{0.73})_4$ nucleated at the high-temperature side of the quartz after cooling down. Energy dispersive spectroscopy (EDS) was applied to determine actual doping rate, where the element ratio of I : Br is 73 : 27 as shown in Fig. S1.

**Structural characterization**. XRD measurements were performed in an AERIS PANalytical X-ray diffractometer at room temperature. The cleaved (001) surface was carefully set to be parallel with the sample stage, and the divergence slit was set at 1/8° with beta-filter Ni accompanied by Cu radiation. For high-angle annular dark-field scanning transmission electron microscopy (HAADF-STEM) measurement of (010) surface, the transmission electron microscopy (TEM) lamella was prepared by using a Zeiss Crossbeam 550 focused-ion-beam (FIB) scanning electron microscopy (SEM). Then, the characterization was conducted on a probe and image corrected FEI Titan Themis Z microscope equipped with a hot-field emission gun working at 300 kV. As shown in Fig. S2, $Bi_4(Br_{0.27}I_{0.73})_4$ has the same stacking structure as $Bi_4Br_4$.

**ARPES measurement**. ARPES measurements were performed on *in-situ*-cleaved thick and homogeneous crystals, attached on the sample holder with torr seal glue with the exposed (001) surface on the sample holder. The helium light (with photon energy ~ 21.2 eV) ARPES characterizations were performed at $T = 7$ K in our laboratory using a DA30L analyzer. The total energy resolution was better than 15 meV, and the angular resolution was set to ~0.3°, which gave a momentum resolution of ~0.01 $\pi/a$.

**Electric transport measurement**. The six-wire Hall-bar-geometry method was used to perform the electric transport measurements. Au was evaporated on freshly cleaved (001) surface of $Bi_4(Br_{0.27}I_{0.73})_4$ at six bonding points to reduce contacting resistance (less than 10 Ω), and silver epoxy was applied to connect the samples with gold wires. The 9 T electric transport data were collected with a physical property measurement system (Quantum Design, Dynacool 9 T) in our home laboratory, where constant D.C. current of 0.5 mA was used to measure resistivity. The high-field electric transport results were measured by pulse magnetic field at Wuhan National High Magnetic Field Center.

**Calculation methods.** The first-principles calculations are performed using the Vienna ab initio simulation package [56] within the projector augmented wave method [57] and the generalized gradient approximation of the Perdew-Burke-Ernzerhof [58] exchange-correlation functional. The energy cutoff of 300 eV is used, and 9×9×3 and 13×13×4 $\Gamma$-centered $k$-grid meshes are adopted for structural relaxation and electronic structure calculations, respectively. The virtual crystal approximation [59] is employed with different ratio for Br and I but the same stacking structure as $Bi_4Br_4$, and the crystal structure of $Bi_4(Br_{1-x}I_x)_4$ is relaxed with van der Waals correction until the residual forces on each atom is less than 0.001 eV/Å. The SOC effect is considered in the electronic structure calculations. Our calculations reveal that a gap closing occurs at the M point for $x = 0.78$ as shown in Fig. S3, indicating an ideal DSM around $Bi_4(Br_{0.22}I_{0.78})_4$, consistent with the experiments. A tight-binding Hamiltonian based on the maximally localized Wannier functions (MLWF) [60] is constructed to further calculate the Berry curvature and anomalous Hall conductivity using the WannierTools package [61]. To simulate the external magnetic field effect, a Zeeman-like manner is employed, where an additional potential of the following form carries the interaction: $V^{\uparrow} = V^{\uparrow} + B_{ext}$ and $V^{\downarrow} = V^{\downarrow} + B_{ext}$ for spin-polarized calculations; $V_{\alpha\beta} = V_{\alpha\beta} + \boldsymbol{B}_{ext} \cdot \sigma_{\alpha\beta}$ with $\sigma$ the vector of Pauli matrices for noncollinear calculations. Although the Landau level effect is not included in the simulations, the basic physical results induced by the Berry curvature and gapless Weyl modes can be definitely determined. First, the Berry curvature is mainly caused by SOC. The bands occupied by integers contribute to saturated Berry curvature and AHC, which is almost unaffected by Landau levels and magnetic field; while the partially occupied bands contribute to non-saturated Berry curvature, and when a Landau band surpassing $E_F$, a change may occur on AHC. Second, the zero-energy mode of gapless Weyl fermions under magnetic field guarantees the existence of the lowest Landau level at $E_F$, ensuring the robustness of chiral anomaly.

**Estimation of effective Landé $g$ factor.** Calculated distances between WP2 $\Delta k_W$ with different effective Zeeman energy $\epsilon_y$ along chain direction are obtained, which are compared with experimental valve of $\Delta k_W$ at critical field of magnetic tunnelling ($B_1$). When experimental and calculated $\Delta k_W$ have the same value of 0.018 Å$^{-1}$, effective Landé $g$ factor is acquired by $g = \frac{\epsilon_y}{B_1 \mu_B} = 88.4$ ($\epsilon_y = 0.11$ eV). The large effective Landé $g$ factor value of 88.4 is similar with that of 77.5 in $Bi_xSb_{1-x}$, [10] coinciding with the fact that large effective Landé $g$ factor value appears with strong SOC strength (provided by heavy element bismuth). Therefore, the value of 88.4 is adopted to estimated calculated magnetic field by $B = \frac{\epsilon}{g\mu_B}$ yielding good consistency with experimental results.

**Data availability:** All data needed to evaluate the conclusions in the paper are present in the paper and available from the corresponding authors upon request. Fig. S10 contains data from published references as described in Supplementary Note 8.

**Code availability:** This study did not involve the development of a custom computer code.

**Acknowledgements**

We are grateful to the Analysis & Testing Center of Beihang University for the facilities and the scientific and technical assistance. This work was supported by the National Natural Science Foundation of China (12574186 J.C.Z., 12274016 Y.D., and 52473287 W.C.H.), the National Key R&D Program of China (2018YFE0202700 Y.D.), and the Fundamental Research Funds for the Central Universities (Grant Nos. 501XSKC2025119001 Y.D., YWF-23SD00-001 J.C.Z. and YWF-22-K-101 Y.D.).

**Author contributions**

J.C.Z. and Y.D. conceived the experiments. J.Y.Z., J.L., A.P.H., Z.X.S. and Z.W.Z. performed the transport measurements. M.Y. carried out ARPES measurements. J.F.W. performed *ab initio* calculations. J.Y.Z. and M.Y. synthesized and characterized the single crystals. Z.Z.R. and Y.S. performed the KPFM measurement. J.Y.Z. wrote the first draft of the paper. J.C.Z., J.F.W., W.C.H., and Y.D. contributed to the revision of the manuscript. J.C.Z. supervised this work. All authors contributed to the scientific planning and discussion.

**Competing interests:** The authors declare no competing interests.

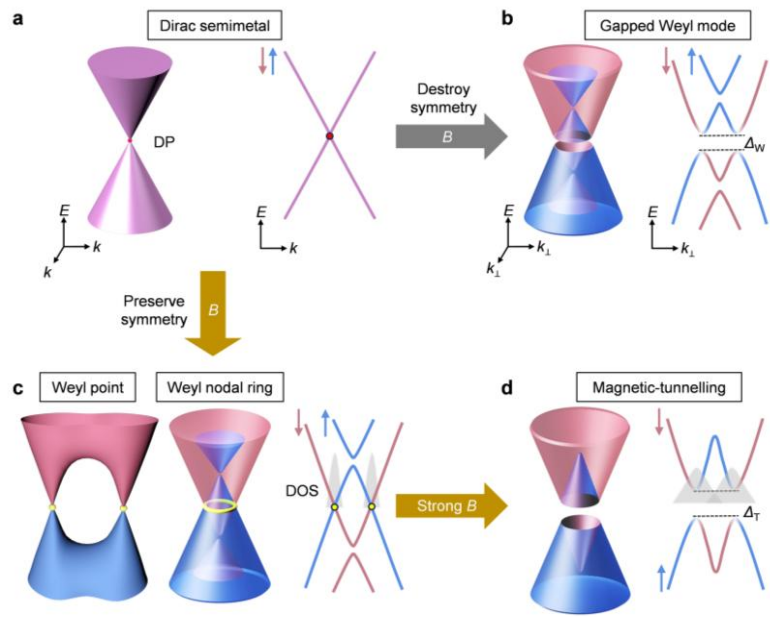

**Fig.1 | Multiple Weyl modes originated from magnetic field manipulation of a Dirac point. a**, Schematics of Dirac point (DP) in a Dirac semimetal with time-reversal symmetry (TRS). Blue and red arrows denote spin-up and spin-down states, respectively. **b**, By applying $B$ to destroy the chiral symmetry, gapped Weyl mode appears with gap $\Delta_W$ due to perturbation of spin-orbital coupling. The $k_\perp$ denotes crystal momentum perpendicular to magnetic field. **c**, Gapless Weyl points (WPs) or Weyl nodal ring (WNR) appear with weak $B$ and preserved symmetry. Grey peaks indicate broadening of density of state (DOS) at WPs or WNR, which is narrow under weak $B$. **d**, Magnetic tunnelling effect and tunnelling gap $\Delta_T$ between WPs induced by strong $B$. Overlapped grey peaks imply coupling between WPs.

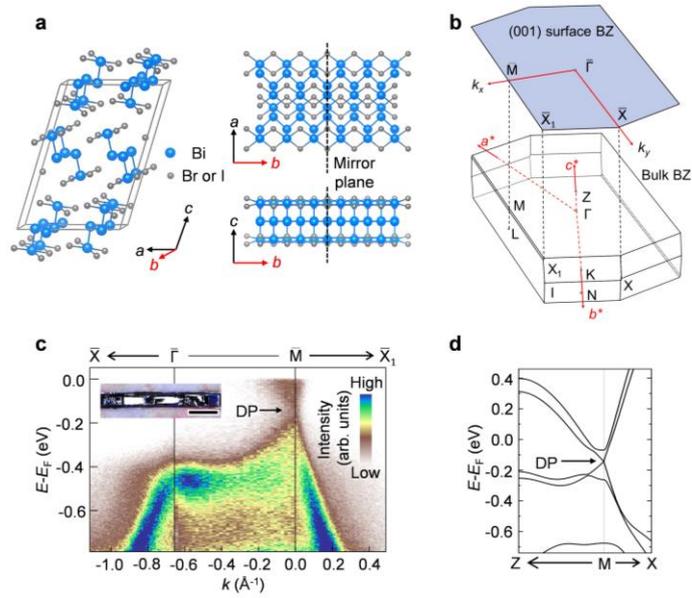

**Fig. 2 | Ideal Dirac semimetal with mirror symmetry. a**, Crystal structure of $Bi_4(Br_{0.27}I_{0.73})_4$, where blue and grey balls represent Bi and Br/I atoms, respectively. Left panel shows the unit cell (grey lines) and quasi-one-dimensional (quasi-1D) chain along *b* axis. Single mirror plane $M_y$ is displayed as dashed lines in right panels. **b**, Bulk Brillouin zone (BZ) and (001) projected BZ with time-reversal invariant momenta (TRIM). **c**, In-plane energy dispersion measured at (001) surface, where a Dirac point (DP) emerges at $\bar{M}$ near 0.14 eV below $E_F$. Inset shows freshly exfoliated (001) surface of $Bi_4(Br_{0.27}I_{0.73})_4$. Scale bar: 0.5 mm. **d**, Calculated bulk band structure with identical dispersion and DP near $M$.

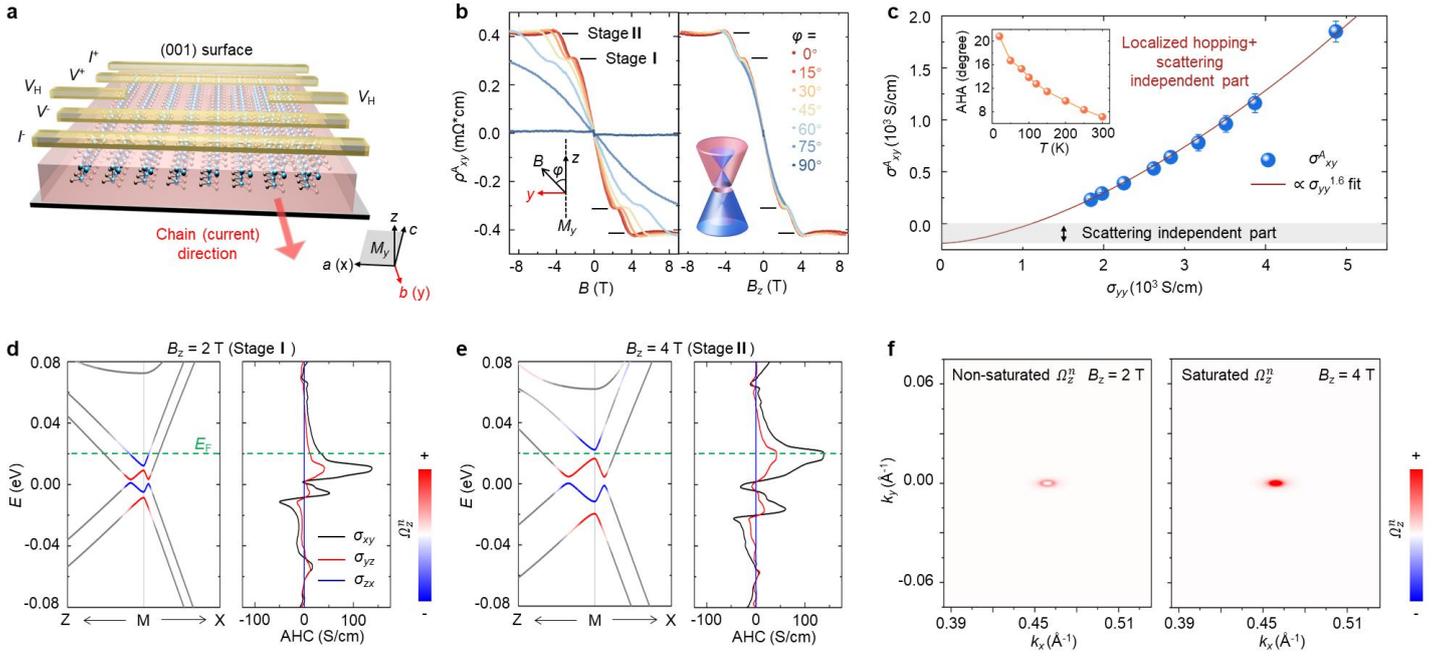

**Fig. 3 | Anomalous Hall effect (AHE) induced by Berry curvature. a**, Configuration of electric transport measurement. Red and yellow cubes denote the sample and electrodes, respectively, where current is applied along chain direction (crystal *b* axis) and Hall voltage is measured along *a* axis. **b**, Anomalous Hall resistivity (AHR) with different angles of magnetic field (left panel) and angle-scaled AHR (right panel). Two stages in AHR are labelled with black lines, and insets define the angle $\varphi$. **c**, Relationship between anomalous Hall conductivity (AHC) and longitudinal conductivity. Dependence of $\propto \sigma_{yy}^{1.6}$ and finite intercept (red line) imply AHE contains both localized hopping part and scattering independent part. The grey region indicates scattering independent contribution of 169 S/cm. Error bars are defined by maximum and minimum valve of AHC as shown in Fig. S9. Inset shows evolution of anomalous Hall angle (AHA) with temperature. Simulated band structure and intrinsic AHC with **d**, $B_z$ = 2 T and **e**, $B_z$ = 4 T respectively. Green dashed line denotes Fermi surface. Red and blue colors represent positive and negative values of Berry curvature $\Omega_z^n$, respectively, which are mainly contributed by the spin-orbital coupling gaps. **f**, Calculated Berry curvature distributions which are summed to Fermi energy under $B_z$ = 2 T (left panel) and $B_z$ = 4 T (right panel).

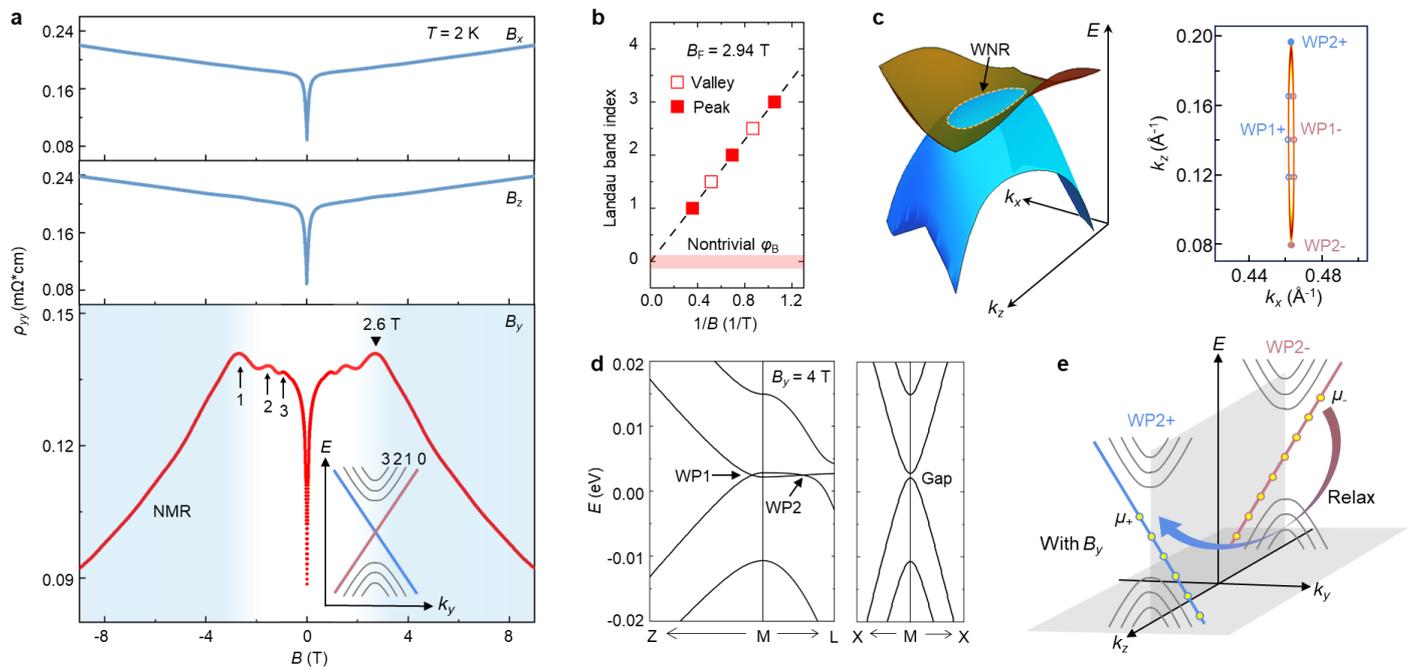

**Fig. 4 | Chiral anomaly in Weyl nodal ring (WNR). a**, Magnetoresistivity (MR) at 2 K with $B$ along $x$ axis (upper panel), $z$ axis (middle panel), and $y$ axis (lower panel). For applying $B_x$ and $B_z$, MR shows monotonically positive tendency with increasing $B$. Only under $B_y$, MR firstly rises accompanied by oscillations (labelled as black arrows) with increasing $B_y$, and then exhibits negative MR (NMR) behavior after the last oscillation. **b**, Shubnikov–de Haas (SdH) oscillation analysis under $B_y$. The solid and hollow squares represent peaks and valleys of oscillation, respectively, acquiring oscillation frequency of $B_F = 2.94$ T. The intercept of linear fit (dashed line) is within ±1/8 (light red region) having nontrivial Berry phase $\varphi_B$. **c**, Left panel: $E$ - $k_x$ - $k_z$ 3D spectra with WNR band crossings denoted by dashed line. Right panel: projected WNR at $k_x$ - $k_z$ plane under $B_y$ = 4 T. Red and blue circles (dots) represent the two different types of Weyl points (WPs) with opposite charity, *i.e.* WP1- (WP2-) and WP1+ (WP2+), respectively. The infinite WPs assemble into the WNR can be disassembled into infinite WP pairs. **d**, Detailed band dispersion around WPs (denoted by black arrows) along $Z$ - $M$ - $L$ path in the $k_y$ = 0 mirror plane (left panel). Calculated band structure along $X$ - $M$ perpendicular to mirror plane shows a gap (right panel). **e**, Schematics of chiral anomaly in WNR under $B_y$. Blue, red, grey lines, and yellow dots represent lowest Landau band with chirality of +1, −1, other higher Landau bands, and occupied states respectively. Electric field parallel to $B_y$ but perpendicular to WP2 separation adds chemical potential difference between two lowest Landau bands ($\mu_- > \mu_+$).

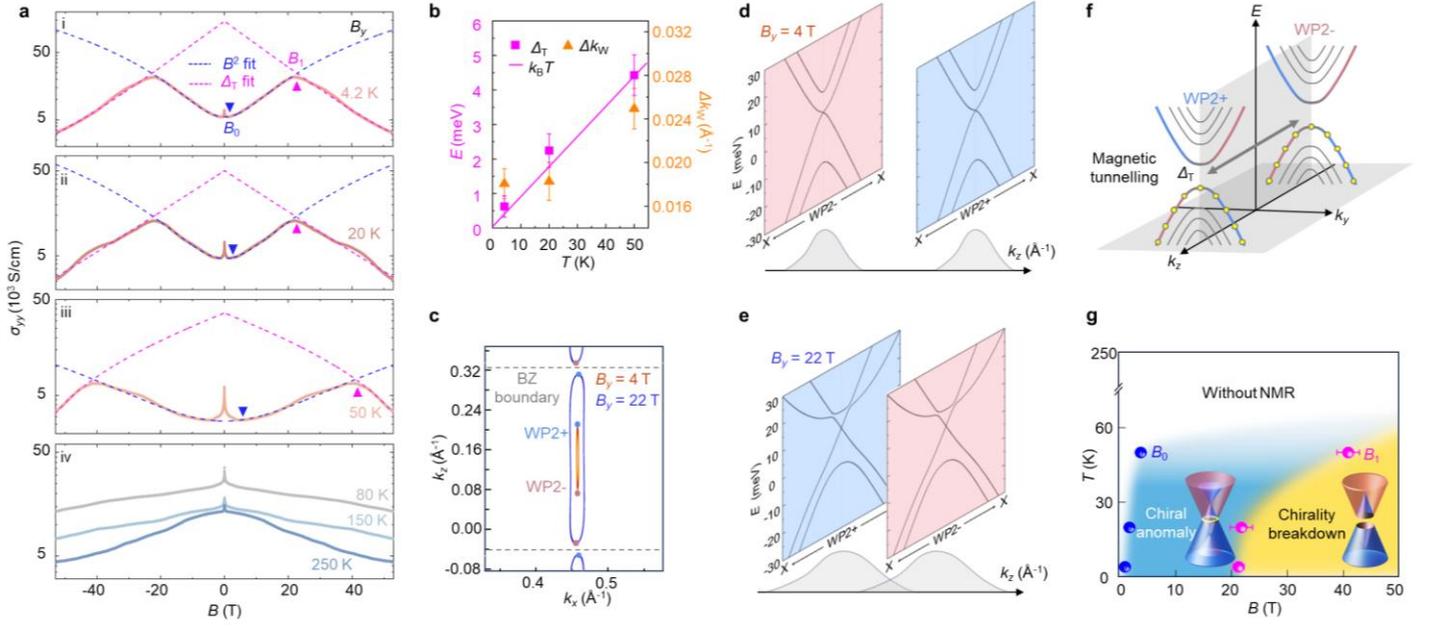

**Fig. 5 | Magnetic tunnelling with chirality breakdown. a**, Magnetoconductivity (MC) under strong $B_y$. For MC at $T$ = 4.2 K, 20 K, and 50 K (panels i to iii), positive MC starts at $B_0$ (blue triangles) and ends at $B_1$ (pink triangles). Blue and pink dashed lines are fitting results of chiral anomaly and magnetic tunnelling, respectively. At higher temperature (panel iv), MC shows monotonically negative behavior. **b**, Fitted magnetic-tunneling gap $\Delta_T$ at $B_1$, which is near the thermal perturbation energy $k_B T$ (pink line). Error bars of $\Delta_T$ are defined by fitted maximum and minimum of Eq. (3). Error bars of $\Delta k_W$ are calculated from $B_1$ by $\sqrt{eB_1/\hbar}$. **c**, Calculated WNR under $B_y$ = 4 T (orange) and $B_y$ = 22 T (blue). Grey dashed lines represent the BZ boundary. **d**, **(e,)** Under $B_y$ = 4 T ($B_y$ = 22 T), distance between WP2- and WP2+ (WP, Weyl point) is larger (smaller) than broadening of their density of state (DOS). Calculated bands along $k_y$ near WP2- and WP2+ are shown in red and blue planes, respectively. Grey humps represent DOS broadening of WPs along $k_z$. **f**, Schematics of magnetic-tunnelling effect between WPs. Overlapped DOS of WPs leads to hybridization of lowest Landau bands, and opens a gap $\Delta_T$ at original location of WPs. **g**, Phase diagram of chiral anomaly (blue) and chirality breakdown (yellow). Insets show the schematical band structure of chiral anomaly and chirality breakdown. NMR, negative magnetoresistivity. Error bars are defined by the magnetic field range of $\sigma_{yy}$ slope change near critical points $B_0$ and $B_1$.